\pdfoutput=1

\documentclass[11pt,twoside,a4paper,cmspaper,final,collab]{cms-tdr}
\def\svnVersion{284525}\def\cmsCernNo{2013-037}\def\cmsCernDate{\today}\def\cmsMessage{Published in the European Physical Journal C as \href{http://dx.doi.org/10.1140/epjc/s10052-015-3367-z}{\doi{10.1140/epjc/s10052-015-3367-z.}}}

\begin{document}\cmsNoteHeader{EXO-12-036}

\hyphenation{had-ron-i-za-tion}
\hyphenation{cal-or-i-me-ter}
\hyphenation{de-vices}
\RCS$Revision: 283324 $
\RCS$HeadURL: svn+ssh://svn.cern.ch/reps/tdr2/papers/EXO-12-036/trunk/EXO-12-036.tex $
\RCS$Id: EXO-12-036.tex 283324 2015-04-03 16:21:40Z chill $
\newlength\cmsFigWidth
\ifthenelse{\boolean{cms@external}}{\setlength\cmsFigWidth{0.98\columnwidth}}{\setlength\cmsFigWidth{0.8\textwidth}}
\ifthenelse{\boolean{cms@external}}{\providecommand{\cmsLeft}{top}}{\providecommand{\cmsLeft}{left}}
\ifthenelse{\boolean{cms@external}}{\providecommand{\cmsRight}{bottom}}{\providecommand{\cmsRight}{right}}
\newlength\cmsFigWidthWider
\ifthenelse{\boolean{cms@external}}{\setlength\cmsFigWidthWider{0.8\textwidth}}{\setlength\cmsFigWidthWider{0.9\textwidth}}

\title{Search for decays of stopped long-lived particles produced in proton-proton collisions at $\sqrt{s}= 8\TeV$}
\titlerunning{Search for decays of stopped long-lived particles produced in proton-proton collisions}
\author{The CMS Collaboration}
\date{\today}

\abstract{
A search has been performed for long-lived particles that could have come to rest within the CMS detector, using the time intervals between LHC beam crossings. The existence of such particles could be deduced from observation of their decays via energy deposits in the CMS calorimeter appearing at times that are well separated from any proton-proton collisions. Using a data set corresponding to an integrated luminosity of 18.6\fbinv of 8\TeV proton-proton collisions, and a search interval corresponding to 281 hours of trigger livetime, 10 events are observed, with a background prediction of $13.2^{+3.6}_{-2.5}$ events. Limits are presented at 95\% confidence level on gluino and top squark production, for over 13 orders of magnitude in the mean proper lifetime of the stopped particle. Assuming a cloud model of R-hadron interactions, a gluino with mass $\lesssim$1000\GeV and a top squark with mass $\lesssim$525\GeV are excluded, for lifetimes between 1\mus and 1000\unit{s}.  These results are the most stringent constraints on stopped particles to date.
}

\hypersetup{%
pdfauthor={CMS Collaboration},%
pdftitle={Search for decays of stopped long-lived particles produced in proton-proton collisions at sqrt(s) = 8 TeV},%
pdfsubject={CMS},%
pdfkeywords={CMS, physics, stopped particle}}

\maketitle

\section{Introduction}
\label{sec:intro}

Many extensions of the standard model (SM) predict the existence of new heavy long-lived particles~\cite{Dimopoulos:1996vz, Baer:1998pg, Jittoh:2005pq, Strassler:2006im, Arvanitaki:2008hq, ArkaniHamed:2004fb}. At the CERN LHC the two general-purpose detectors, ATLAS and CMS, have already set stringent limits on the existence of such particles with searches that exploit the anomalously large ionization and/or long time-of-flight that they would exhibit as they traverse the detectors~\cite{Chatrchyan:2013oca, Aad:2012pra}. These searches are complemented by those that target the fraction of such particles produced with sufficiently low kinetic energy (KE) that they come to rest in the detectors. In this approach, the subsequent decay is directly observed, allowing (in principle) the reconstruction of the ``stopped" particle and the study of its characteristics~\cite{Graham:2011ah}.

New long-lived heavy particles, such as heavy gluinos (\PSg) and top squarks (\PSQt), could be pair-produced in proton-proton (pp) collisions and combine with SM particles to form R-hadrons~\cite{Fayet:1977yc,Fayet:1978qc,Farrar:1978xj}. These R-hadrons would then traverse the volume of the detector, interacting with detector materials via nuclear interactions and, if charged, by ionization. Below a critical velocity ${\lesssim}0.45c$, the KE of the R-hadron is small enough and the energy loss per unit length ($\rd{}E/\rd{}x$) large enough that the particle can come to rest within the body of the detector. At some later time, the stopped R-hadron would then decay. Assuming at least one daughter particle is a SM particle and the R-hadron has stopped in an instrumented region of the detector, the decay could be observable. If this stopping location is in the calorimeter, as is most likely given its density, the experimental signature would be a randomly-timed, relatively large energy response spread over a few channels. Since these depositions might be difficult to differentiate from those of SM particles produced in pp collisions, they would be most easily observed at times between pp collisions.  During these times the detector should be quiet with the exception of cosmic rays, some beam-related backgrounds, and instrumental noise. The results of such searches have previously been reported by the D0 collaboration at the Tevatron~\cite{Abazov:2007ht}, and by the CMS~\cite{Chatrchyan:2012dxa,Khachatryan:2010uf} and ATLAS collaborations~\cite{Aad:2012zn,Aad:2013gva}.

This paper provides an update to the CMS search for stopped particles. The new analysis benefits from a four-fold integrated luminosity increase and uses data resulting from higher energy pp collisions compared to the previous CMS publication~\cite{Chatrchyan:2012dxa}.

\section{The CMS detector and jet reconstruction}
{\tolerance=600
The central feature of the CMS apparatus is a superconducting solenoid of 6\unit{m} internal diameter, providing a magnetic field of 3.8\unit{T}. Within the superconducting solenoid volume are a silicon pixel and strip tracker, a lead tungstate crystal electromagnetic calorimeter (ECAL), and a brass/scintillator hadron calorimeter (HCAL), each composed of a central (barrel) and two forward (endcap) sections. In the region $\abs{ \eta }< 1.74$, the HCAL cells have widths of 0.087 in pseudorapidity and 0.087 in azimuth ($\phi$). In the $\eta$-$\phi$ plane, and for $\abs{\eta}< 1.48$, the HCAL cells map onto $5 \times 5$ ECAL crystals arrays to form calorimeter towers projecting radially outwards from close to the nominal interaction point. At larger values of $\abs{ \eta }$, the size of the towers increases and the matching ECAL arrays contain fewer crystals. Within each tower, the energy deposits in ECAL and HCAL cells are summed to define the calorimeter tower energies, subsequently used to provide the energies and directions of hadronic jets.  Muons are measured in gas-ionization detectors embedded in the steel flux-return yoke outside the solenoid; drift tubes and resistive-plate chambers (RPC) provide coverage in the barrel, while cathode strip chambers (CSC) and RPC provide coverage in the endcaps. Extensive forward calorimetry complements the coverage provided by the barrel and endcap detectors. The first level (L1) of the CMS trigger system, composed of custom hardware processors, uses information from the calorimeters and muon detectors to select the most interesting events in a fixed time interval of less than 4\mus. The high-level trigger (HLT) processor farm further decreases the event rate from around 100\unit{kHz} to around 400\unit{Hz}, before data storage. A more detailed description of the CMS detector, together with a definition of the coordinate system used and the relevant kinematic variables, can be found in Ref.~\cite{Chatrchyan:2008zzk}.
\par}

Because a stopped particle is by definition at rest, the energy deposits in the calorimeter that result from its decay would not generally be oriented in towers radially towards the pp interaction point of CMS. Nevertheless, such depositions are sufficiently jet-like that they may be reconstructed offline using the anti-\kt clustering algorithm~\cite{Cacciari:2008gp, Cacciari:2011ma} with a distance parameter of 0.5.

\section{Data set and Monte Carlo simulation samples}
\label{sec:datasets}

{\tolerance=400
The search is performed using $\sqrt{s} = 8\TeV$ pp collision data collected between May and December 2012, corresponding to an integrated luminosity of 18.6\fbinv and to 281 hours when the dedicated trigger used in this analysis was active (``livetime"). The maximum instantaneous luminosity achieved during this period was 7.5$\times 10^{33}\cm^{-2}\unit{s}^{-1}$. As a control sample, this analysis uses $\sqrt{s} = 7\TeV$ pp collision data corresponding to an integrated luminosity of 3.6\pbinv collected at the beginning of LHC operations in 2010. The control sample includes 253 hours of trigger livetime. Though the integrated luminosity for this period is much smaller than that in 2012, the trigger livetime is comparable because of the longer time interval between collisions in 2010.
\par}

Simulated signal Monte Carlo (MC) events for this analysis are generated in three stages. In the first stage we use \PYTHIA 8.153~\cite{Sjostrand:2007gs} to generate $\Pp\Pp\to \PSg\PSg$ and $\Pp\Pp\to \PSQt\PASQt$ events. The colored sparticles are hadronized with the default parameters in the \textsc{Rhadrons} package included in \PYTHIA 8.153. These parameters influence technical aspects of the hadronization process, e.g. the fraction of produced R-hadrons that contain a gluino and a valence gluon, which is set to 10\%. Because of the nature of the stopped-particle technique, these parameters do not have a significant effect on the phenomenology. The passage of the R-hadrons through the detector is simulated with \GEANTfour 9.4.p03~\cite{Agostinelli:2002hh}. A phase-space driven ``cloud model'' of R-hadron interactions with the material of the CMS detector~\cite{Kraan:2004tz,Mackeprang:2006gx}, referred to as the ``generic'' model in Ref.~\cite{Aad:2012zn}, is used to simulate the interaction of these R-hadrons with the CMS detector. In this model, which has emerged as the standard benchmark for these searches, R-hadrons are treated as supersymmetric particles surrounded by a cloud of loosely bound quarks or gluons. 

The KE of these simulated R-hadrons is diminished though nuclear interactions and ionization, and they can thus come to rest within the body of the CMS detector. If this occurs, the position in the CMS coordinate system of the stopped R-hadron is recorded. The second stage of the simulation generates an R-hadron, translates it to the stopping position recorded in the first stage, and causes it to decay at rest via a second \GEANTfour step. The gluino decay is simulated as $\PSg\to \Pg\PSGcz$ and the top squark decay is simulated as $\PSQt\to \PQt\PSGcz$, where $\PSGcz$ is the lightest neutralino in both instances. The first stage allows estimation of the stopping probability $\varepsilon_\text{stopping}$, and the second stage allows estimation of the reconstruction efficiency $\varepsilon_\text{reco}$. While any spin correlation of the decaying gluino with its pair produced partner is lost in this approach, no observable effects of this omission are expected~\cite{Kramer:2009kp}.  The probability for the subsequent decay of a stopped particle to occur at a time when the trigger is live is estimated with a custom third stage pseudo-experiment MC simulation. We randomly generate decay times according to the exponential distribution expected for a given lifetime hypothesis and compare these to the delivered luminosity profile and actual bunch structure of each LHC fill, and all relevant CMS trigger rules, in order to determine an effective luminosity ($L_\text{eff}$) for each lifetime hypothesis.

\section{Event selection}
\label{sec:selection}

We perform the search with a dedicated trigger used to select events out-of-time with respect to collisions. A series of offline selection criteria are then applied to exclude events likely due to background processes.

The LHC beams are composed of circulating bunches of protons. At the experiments, bunch crossings (BX) nominally occur at intervals of 25\unit{ns}. In 2012, however, the LHC operated with a 50\unit{ns} minimum interval between proton bunches and the most often used LHC filling scheme had 1377 of such intervals containing colliding bunches of protons. This search looks for events in a suitable subset of the other 802 BXs i.e. during intervals that are ``out-of-time" with respect to normal collisions. Such events are recorded using an updated version of the calorimeter trigger employed in the earlier CMS study~\cite{Chatrchyan:2012dxa,Khachatryan:2010uf} that uses the two beam position and timing monitors (BPTX) that are positioned along the beam axis, at either end of the CMS detector close to the beams. These BPTX are sensitive electrostatic instruments that are able to detect the passage of an LHC proton bunch. Consequently, in order to search for out-of-time events, we employ a dedicated trigger that requires an energy deposit in the calorimeter trigger together with the condition that neither BPTX detects a bunch in that BX. We also require that at most one BPTX produces a signal in a window $\pm$1~BX around the triggered event. This rejects triggers due to out of time ``satellite'' bunches that occasionally accompany the colliding protons. The energy deposition in the calorimeter of a jet from the R-hadron decay is sufficiently similar to those of jets originating directly from pp collisions that a calorimeter jet trigger can be used.  At L1 the jet transverse energy, which is calculated assuming the jet was produced at the nominal interaction point, is required to be greater than 32\GeV, while in the HLT jet energy is required to be greater than 50\GeV. At both L1 and HLT $\abs{\eta_\text{jet}}$ is required to be less than 3.0. Finally, the trigger rejects any event within a $\pm$1~BX window that is identified as beam halo at L1 by the presence of a pair of CSC hits geometrically consistent with the expected trajectory of a halo muon.

We select events more than $\pm$1 BX away from an in-time pp collision. Additionally, to remove rare events in which an out-of-time pp collision occurred, usually caused by residual protons found in between proton bunches, we veto events that include a primary vertex reconstructed by an adaptive vertex fit~\cite{Chatrchyan:2014fea} with greater than or equal to 4 degrees of freedom as expected from a pp collision. Finally, we require that a jet is reconstructed with an energy of at least 70\GeV. This threshold is set just above the turn-on plateau for the 50\GeV trigger.

Halo muons are a source of background for this analysis. Halo muons are produced when off-orbit protons in the LHC beam strike material in some limiting aperture of the LHC upstream of the CMS detector. The resulting collision produces a shower of particles, most of which decay before reaching CMS. Muons are produced in these decays and, given their long lifetimes and the fact that they undergo only electroweak interactions, can survive long enough to traverse the detector. When they pass through the denser regions, they can emit a bremsstrahlung photon that strikes the calorimeter and can be reconstructed with large enough energy to be included in the search sample. We remove these events by vetoing any event in which there are hits recorded within the CSC forward muon chambers. The low-noise rate of the CSC detectors allows the requirement to be set at the single-hit level, which enables the maximal exclusion of this background.

Muons from cosmic rays incident on the CMS detector can also mimic the signal characteristics. Similar to the halo background, cosmic ray muons may emit a photon that strikes the calorimeter, leaving a large energy deposit. To remove such events, we consider the distribution of reconstructed hits within the barrel muon system. Compared to the expected signal, there are key differences with cosmic ray muons that can be exploited. It is possible that heavy R-hadron decay products have a large enough energy to ``punch through" the outer region of the calorimeter and the first layers of the iron yoke of the solenoid, leaving energy deposits in the muon system. This phenomenon is easily distinguished from cosmic ray muons by considering the distribution of reconstructed hits. In the case of cosmic ray muons, we expect hits evenly distributed throughout the barrel of the muon system, whereas for signal events, we expect the hits to be restricted to the innermost layers of the muon system. Additionally, in the case of punch-through, the muon hits should be localized near the reconstructed jet. Unlike signal events, hits from cosmic ray muons may also appear opposite in $\phi$ to that of the reconstructed jet. Exploiting these properties, we are able to substantially reduce the cosmic ray background contaminating the signal region by removing events with hits in the outer layers of the muon system ($r > 560$~cm), events with hits recorded in both the top and bottom of the muon system, and events with hits in the muon system separated from the leading jet in $\phi$ by greater than 1.0 radians.

The final source of background stems from instrumental noise in the calorimetry system, primarily within the HCAL. Noise in the HCAL can give rise to events in which an errant spike in energy is recorded, unrelated to any physical interaction with particles produced inside the detector. These occurrences are rare, but the calorimeter response resembles the anticipated signal and must be removed. The HCAL electronics has a well-defined time response to charge deposits generated by showering particles. Analog signal pulses produced by the electronics are sampled at 40~MHz, synchronized with the LHC clock. These pulses are readout over ten BX samples centered around the pulse maximum. A physical signal exhibits a clear peak followed by an exponential decay over the next few BXs. Moreover, energy deposits from physical particles tend to have a large fraction of the pulse energy in the peak BX. We use a series of offline criteria that exploit these timing and topological characteristics to remove spurious events due to noise, which do not exhibit these properties. These criteria are detailed in Ref.~\cite{Khachatryan:2010uf} and are applied as in that analysis with the exception of the requirement that the leading jet has at least 60\% of its energy contained in fewer than 6 towers.

\section{Signal efficiency}
\label{sec:sigeff}

Using the first stage of the MC simulation described in Section~\ref{sec:datasets}, we estimate the probability of an R-hadron to stop within the instrumented regions of the detector. In particular, we are interested in R-hadrons that stop in the barrel region of the calorimeter, since these are the regions where we can observe the subsequent jet-like energy deposits from the decay products. We exclude the endcap calorimeters since the signal-to-background ratio is less favorable in this region. R-hadrons could also stop within the iron yokes interleaved with the muon detector system, but we expect a negligible efficiency to detect the corresponding decays. The simulation demonstrates that the stopping probability is approximately constant over the range of R-hadron masses considered in this search. The probability that at least one R-hadron is stopped within the barrel region of the calorimeter is found to be 8\% for gluinos with $m_{\PSg}= 800\GeV$ and 6\% for top squarks with $m_{\PSQt} = 400\GeV$. The slighter larger $\varepsilon_\text{stopping}$ obtained for gluinos is because of their greater propensity to form doubly charged states.

\begin{figure}[th]
 \centering
   \includegraphics[width=\cmsFigWidth]{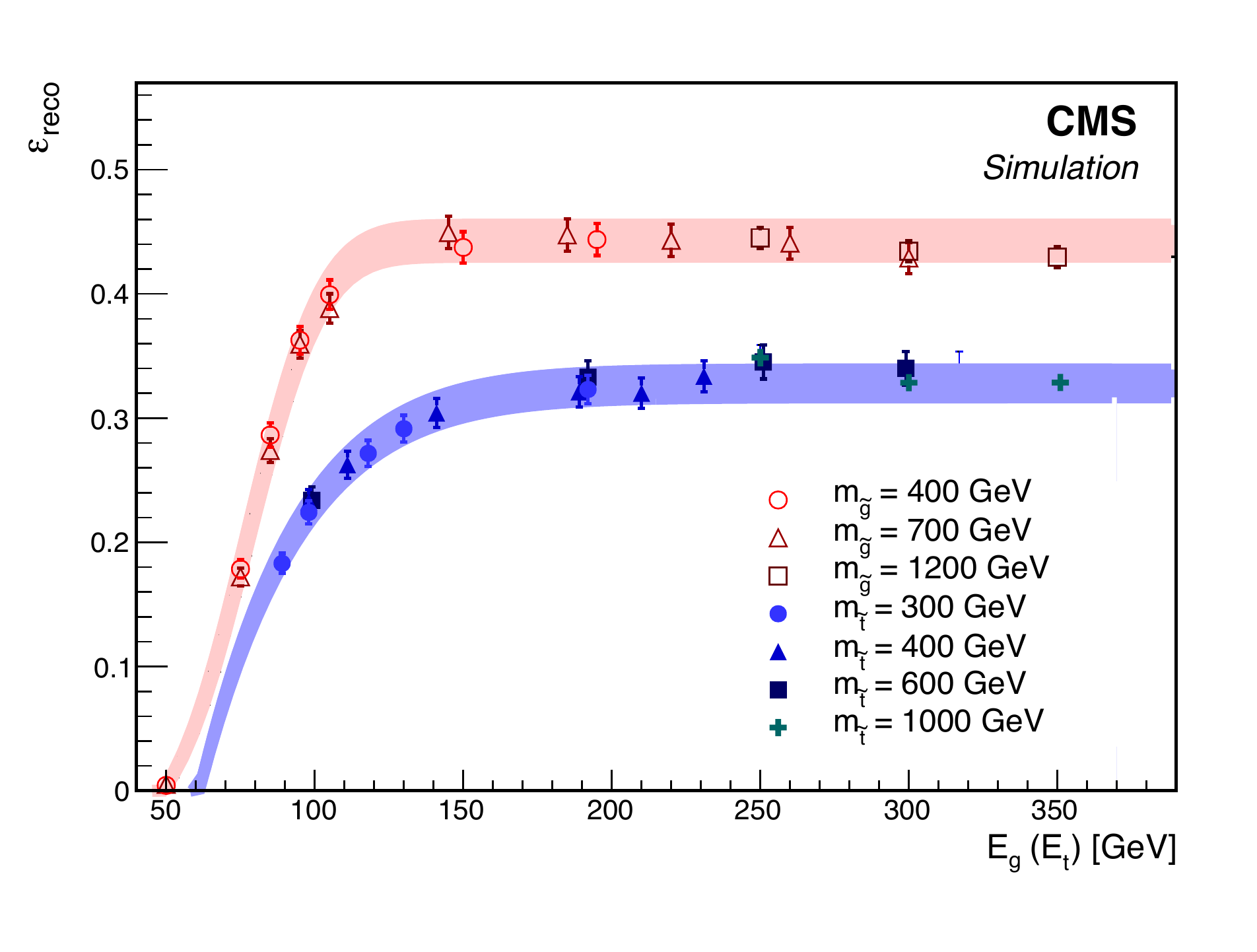}
   \caption{The reconstruction efficiency $\varepsilon_\text{reco}$ for \PSg{} and \PSQt R-hadrons that stop in the barrel region of the calorimeter as a function of the energy of the produced SM daughter particle. The shaded bands indicate the systematic uncertainty in $\varepsilon_\text{reco}$.}
   \label{fig:deteff}
\end{figure}

For the particles that stop and then decay within the calorimeter, MC simulations estimate an approximate trigger efficiency of 70\%. We define $\varepsilon_\text{reco}$ as the number of signal events that pass all selection criteria (including the trigger requirement) divided by the number of signal events that stop within the barrel region of the calorimeter. The reconstruction efficiency depends principally on the energy of the visible daughter particle of the R-hadron decay, which we denote by $E_{\Pg}$ ($E_{\PQt}$) if the daughter is a gluon (top quark). The reconstruction efficiencies obtained for gluinos and top squarks are plotted as a function of this energy in Fig.~\ref{fig:deteff}. Above the minimum energy threshold for the SM decay products, where $\varepsilon_\text{reco}$ becomes approximately constant, $E_{\Pg} > 120\GeV$ ($E_{\PQt} > 150\GeV$), we obtain $\varepsilon_\text{reco} \approx 45\%$ (32\%) for $\PSg (\PSQt$) decays. The top squark efficiency is lower than the gluino efficiency primarily because of $\PQt \to \PQb\Pgm\Pgn$ decays that yield less visible energy in the calorimeter and are rejected by the muon vetoes.  When $E_{\PQt}$ is below $m_{\PQt}$, which can happen in cases when the mass splitting between the \PSQt and $\PSGcz$ is small, the top quark is off the mass-shell.

The signal efficiency is given by the product of $\varepsilon_\text{stopping}$ and $\varepsilon_\text{reco}$.

\section{Backgrounds}
\label{sec:bkg}

It is possible for halo muons to escape detection in the endcap muon system. Escaping detection is uncommon, but owing to the high rate of halo production in the 2012 data collection period, the expected halo background is non-negligible. We estimate the halo veto inefficiency using a ``tag-and-probe" method~\cite{firstWZpaper} that analyzes a high-purity sample of halo muons to determine the rates at which we record hits on both ends of the endcap muon detectors, compared to the rate at which we see only the ``incoming" or  ``outgoing" portions of the halo muon track. Because of timing and trigger effects, we may only observe the outgoing leg of the halo muon, with the incoming leg recorded in a previous BX. When the reverse occurs, we see only the incoming leg. Additionally, it is possible for the muon to lose all of its energy within the CMS detector before reaching the opposite side CSC chambers, which also results in an incoming-only event. We classify these events as to whether the halo originates from the clockwise or counterclockwise beam, and bin them by their geometric location in the endcap muon system. After integrating these distributions, we measure a halo veto inefficiency of $1\times10^{-5}$. This inefficiency is multiplied by the number of halo events vetoed from the search sample yielding an average halo background estimate of $8.0 \pm 0.4$ events, where the uncertainty in the estimate is owing to the limited size of the data samples used.

To determine the rate at which cosmic ray muons escape detection by the cosmic muon veto, we generate a sample of 300 million simulated cosmic events using \textsc{Cmscgen}~\cite{Biallass:2009ev}, which is based on the air shower program \textsc{Corsika}~\cite{Heck:1998vt}, and has been validated in~\cite{Khachatryan:2010mw}. After requiring a substantial energy deposit in the calorimeter and the absence of any hits in the muon endcap system that would cause the event to be classified as a halo muon, we estimate the inefficiency of the cosmic muon veto by dividing the number of events that escape this veto by the total number of events. The inefficiency obtained in this manner is roughly 0.5\%. After multiplication by the number of cosmic ray muon events vetoed from the search sample, this corresponds to a predicted cosmic background of $5.2 \pm 2.5$ events, where the uncertainty in the estimate is owing to the limited size of the data samples used.

\begin{figure}[th]
 \centering
 \includegraphics[width=\cmsFigWidth]{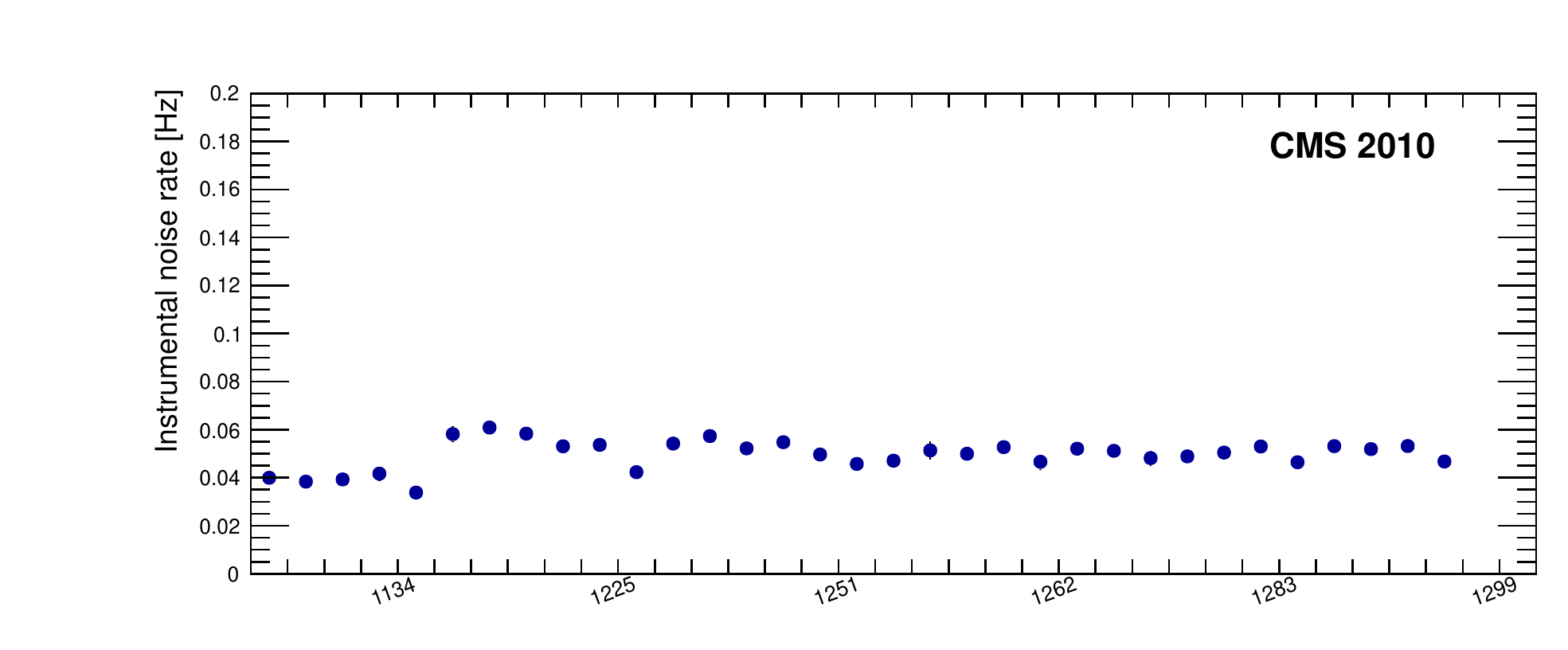}
 \includegraphics[width=\cmsFigWidth]{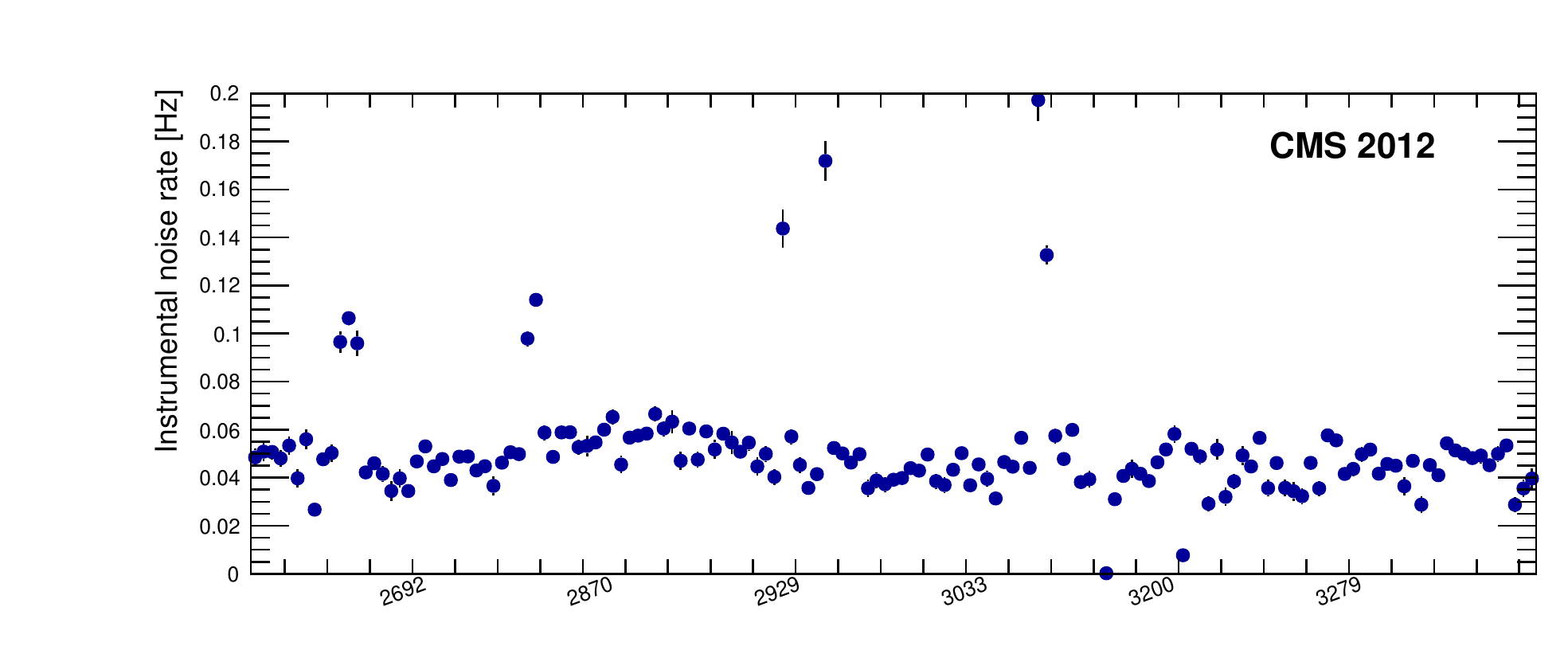}
   \caption{The noise rates from 2010 data (top) and 2012 data (bottom).}
   \label{fig:noiserates}
\end{figure}

Finally, the background owing to instrumental noise is estimated by considering data recorded in 2010. Figure~\ref{fig:noiserates} shows the measured noise rates in both periods.  In these plots all selection criteria except those that are designed to reject noise are applied. Additionally, only events at least 5 BX from a bunch are considered. This is done to reduce halo contamination in the distributions, which is abundant directly after a bunch crossing. There is a greater variation in the 2012 noise rate because of increased halo background, which can mimic HCAL noise if no CSC hits are present. The small variation seen in the 2012 data, while larger than that seen in 2010, is nevertheless small compared to the systematic uncertainty in the noise event count. The larger variations observed in the 2012 data are attributed to residual halo contamination arising from non-standard LHC beam conditions.

Because the rate of instrumental noise is approximately constant between the two periods, and the data recorded in early 2010 were delivered with very low instantaneous luminosity, this sample is unlikely to contain either halo events or signal events. After applying the same selection criteria to the 2010 data sample as used in this analysis, two events remain. We estimate the cosmic ray muon contribution to this sample to be $4.8 \pm 3.6$ events. Because the cosmic ray background estimate exceeds the number of observed events, we assume a central value of zero events for the instrumental noise contribution to the 2010 sample, and allowing for Poisson fluctuations in both the noise and cosmic ray muon contributions, set a 68\% confidence level (CL) upper limit on this contribution of 2.3 events. This estimate is then scaled by the ratio of the 2012 and 2010 livetimes, resulting in an expected noise contribution of $0.0^{+2.6}_{-0.0}$ events in the 2012 data set.

\section{Sources of systematic uncertainties}
\label{sec:systematics}

The model-independent results of the counting experiment described in this paper have relatively few systematic uncertainties. There is a 2.5\% uncertainty in the integrated luminosity~\cite{LUM-13-001}. There is a 13\% uncertainty in the reconstruction efficiency resulting from the possibility that even above the minimum value of $E_{\Pg}$ ($E_{\PQt}$), this efficiency is not completely independent of the energy of the daughter particle as is assumed.  This uncertainty is determined by considering the difference between the individual values of $\varepsilon_\text{reco}$ in Fig.~\ref{fig:deteff} and the average value for all points above the minimum value of $E_{\Pg}$ ($E_{\PQt}$). The shaded bands in the figure depict this uncertainty. Because the energy deposits resulting in reconstructed jets are not the result of jets originating from the center of the detector (as is the case for jets originating from pp collisions), they are not necessarily directed radially, and standard uncertainties in the jet energy scale (JES) do not apply. Instead, we determine the JES uncertainty by referencing a study performed on the HCAL during cosmic ray data taking in 2008~\cite{HCALperformance}. This study compares the reconstructed energy deposits in the HCAL for simulated cosmic ray events and cosmic ray events in 2008 data.  These comparisons lead to an estimated uncertainty of $\sim$2\% on the simulation. A similar study comparing data and simulation for dijets originating at the interaction point conducted with 2012 data yielded an uncertainty of $<$2\% for jets striking the HCAL barrel with angles of incidence from 0 to 60 degrees~\cite{FitzgeraldSchroeder:1702650}. While the study demonstrates that the HCAL response is well simulated with an uncertainty of about 1\%, we take a conservative JES uncertainty of 3\% to compensate for any effects of stopped particle decays that these studies cannot test because of the potentially large angles that could sometimes be expected in the signal decays. This value for the JES uncertainty leads to an uncertainty in the search results of about 2\% at the minimum value of $E_{\Pg}$. The value of 3\% is somewhat pessimistic since the uncertainty falls rapidly as $E_{\Pg}$ increases.  Variations in the reconstructed jet energy are only important for deposits with energies close to the jet energy threshold, which typically correspond to events in which $E_{\Pg}$ is small.

In obtaining constraints on a particular model, however, more substantial uncertainties arise since the signal yield is sensitive to the stopping probability. While the \GEANTfour simulation used to derive the stopping probability very accurately models both the electromagnetic and nuclear interaction energy-loss mechanisms, the relative contributions of these energy-loss mechanisms to the stopping probability depends significantly on unknown R-hadron spectroscopy. We do not consider this dependence to be a source of error, however, since given a particular model for the spectrum the resultant uncertainty in the stopping probability is small.

In addition to these uncertainties in the signal efficiency, there is also a systematic uncertainty in the background estimate described in Section~\ref{sec:bkg}. This systematic uncertainty arises from the limited size of the data control samples that were used to estimate the contribution of each of the background processes to the search sample.

The systematic uncertainties are summarized in Table~\ref{tab:systematics}.

\begin{table}[thb]
  \topcaption{Summary of systematic uncertainties.}
  \label{tab:systematics}
  \centering
  \begin{tabular}{lc} \hline
  	Systematic uncertainty 	& Fractional uncertainty \\ \hline
	JES uncertainty		& $\pm$3\% \\
	Luminosity uncertainty	& $\pm$2.5\% \\
	$\varepsilon_\text{reco}$ uncertainty	& $\pm$13\% \\
	Background uncertainty 	& $+27$\%, $-19$\% \\ \hline
  \end{tabular}

\end{table}

\section{Results}
\label{sec:results}
The total and individual background estimates for both the 2012 search period and the 2010 control period used to determine the background from instrumental noise, are summarized in Table~\ref{tab:results}, together with the number of observed events.

\begin{table*}[thb]
  \topcaption{Background predictions and observed events for the 2010 control and 2012 search samples.}
  \label{tab:results}
  \centering
    \renewcommand{\arraystretch}{1.25}
  \begin{tabular}{  ccccccc} \hline
  	Period	&Trigger livetime~(h) &$N^\text{bkg}_\text{noise}$	&$N^\text{bkg}_\text{cosmic}$	&$N^\text{bkg}_\text{halo}$		&$N^\text{bkg}_\text{total}$ 			&$N^\text{obs}$ \\ \hline
	2010		&253	&$0.0^{+2.3}_{-0.0}$	&$4.8 \pm 3.6$	&---	&$4.8^{+4.3}_{-3.6}$	& 2 \\
	2012		&281	&$0.0^{+2.6}_{-0.0}$	&$5.2 \pm 2.5$		&$8.0 \pm 0.4$	&$13.2^{+3.6}_{-2.5}$	&10 \\ \hline
\end{tabular}

\end{table*}

With the assumption that the backgrounds listed in Table~\ref{tab:results} are uniformly distributed in time, which is valid even for the halo background for times at least one BX away from the collision as our selection requires, we perform a counting experiment in equally spaced $\log$(time) bins of gluino (top squark) lifetime hypotheses, $\tau_{\PSg} (\tau_{\PSQt})$, from $10^{-7}$  to $10^6$ seconds. For lifetime hypotheses shorter than one orbit (89\mus), we count only candidates within a sensitivity-optimized time window of $1.3 \tau_{\PSg}( \tau_{\PSQt})$ from any pp collision. This restriction avoids the addition of backgrounds for time intervals during which the signal has a high probability to have already decayed. In order to resolve any time structure in the data within a single orbit, we test two additional lifetime hypotheses for each observed event for these counting experiments:  the largest lifetime hypothesis for which the event lies outside $1.3 \tau_{\PSg}( \tau_{\PSQt})$, and the smallest lifetime hypothesis for which the event is contained within $1.3 \tau_{\PSg}( \tau_{\PSQt})$. Table~\ref{tab:results2} shows the results of the counting experiments for selected lifetime hypotheses.  The observed number of events is consistent with the background expectation for all lifetime hypotheses tested.

\begin{table*}[thb]
  \topcaption{Results of counting experiments for selected lifetime hypotheses.}
  \label{tab:results2}
  \centering
  \renewcommand{\arraystretch}{1.25}
  \begin{tabular}{ccccc} \hline
	Lifetime hypothesis		&$L_\text{eff}$ (fb$^{-1}$)	&Trigger livetime (s)		&Expected bkg.		&Observed \\ \hline
	50\unit{ns}			&0.121		&$5.0\times10^4$	&$0.66^{+0.18}_{-0.07}$	&0 \\
	75\unit{ns}			&0.271		&$1.0\times10^5$	&$1.3^{+0.4}_{-0.2}$		&3 \\
	100\unit{ns}			&0.512		&$2.0\times10^5$	&2.6$^{+0.7}_{-0.5}$		&3 \\
	1$\mus$			&2.864		&$8.4\times10^5$	&11.0$^{+3.0}_{-2.1}$	&6 \\
	10$\mus$		&3.885		&$1.0\times10^6$	&13.1$^{+3.6}_{-2.4}$	&10 \\
	100$\mus$		&3.972		&$1.0\times10^6$	&13.2$^{+3.6}_{-2.5}$	&10 \\
	$10^3$\unit{s}		&3.868		&$1.0\times10^6$	&$13.2^{+3.6}_{-2.5}$	&10 \\
	$10^4$\unit{s}		&3.004		&$1.0\times10^6$	&$13.2^{+3.6}_{-2.5}$	&10 \\
	$10^5$\unit{s}		&1.727		&$1.0\times10^6$	&$13.2^{+3.6}_{-2.5}$	&10 \\
	$10^6$\unit{s}		&1.181		&$1.0\times10^6$	&$13.2^{+3.6}_{-2.5}$	&10 \\ \hline
  \end{tabular}

\end{table*}

\subsection{Limits on gluino and top squark production}
\begin{figure*}[th]
 \centering
   \includegraphics[width=\cmsFigWidthWider]{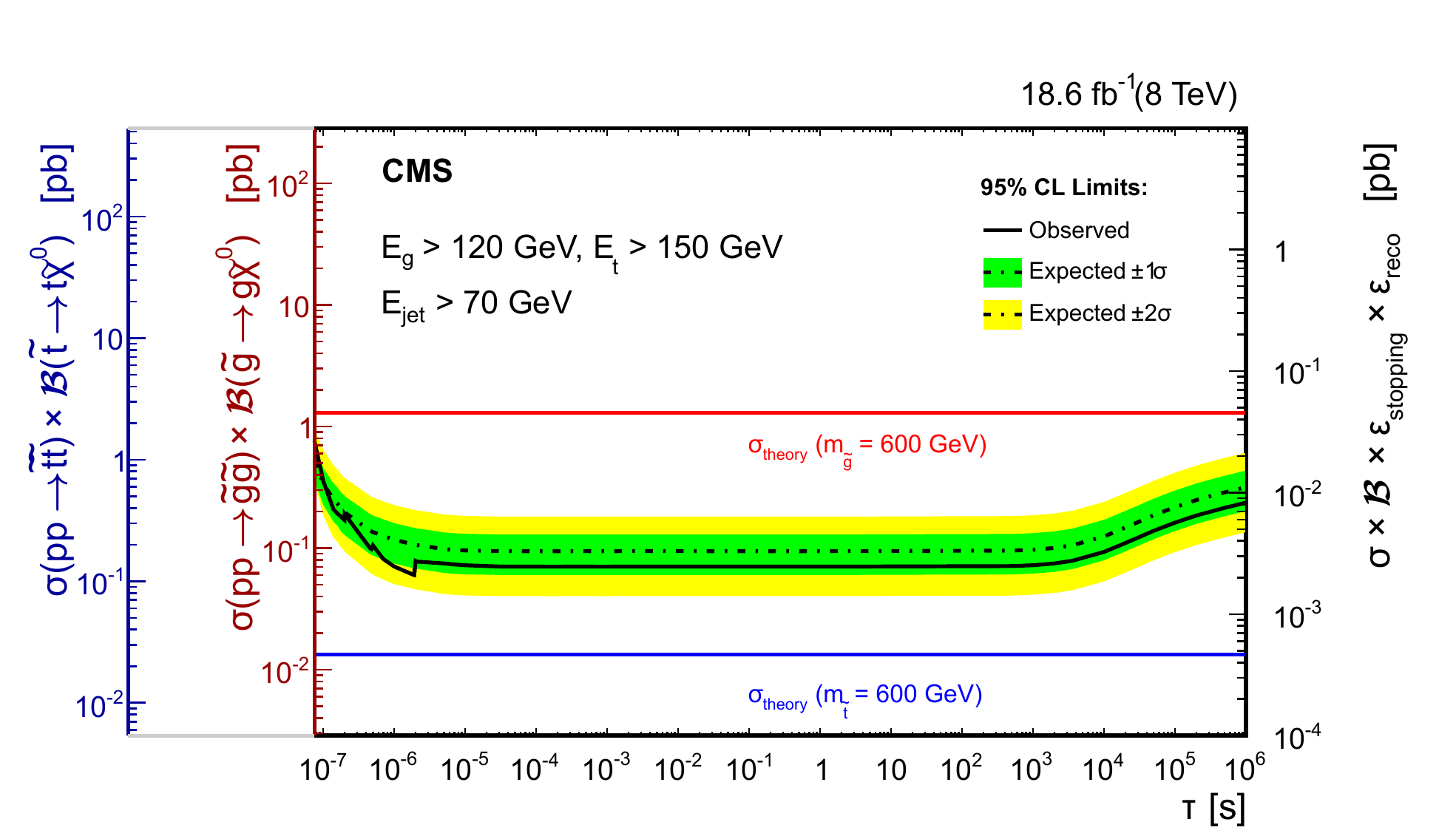}
   \caption{The left-hand axes present expected and observed 95\% CL upper limits on top squark and gluino pair production cross sections using the cloud model of R-hadron interactions, as a function of particle lifetime. The NLO+NLL cross sections shown were obtained with \textsc{NLL-Fast}~\cite{Beenakker:2011fu}. The right-hand axis shows the quantity   $\sigma\times\mathcal{B}\times\varepsilon_\text{stopping}\times  \varepsilon_\text{reco}$, which is more model independent. The structure observed between $10^{-7}$ and $10^{-5}$\unit{s} is due to the number of observed events incrementing when crossing boundaries between lifetime bins. When $E_{\PQt} < m_{\PQt}$, the top quark is off the mass shell.}
   \label{fig:70gevxsec}
\end{figure*}

We obtain upper limits on the signal production cross section using a hybrid CL$_\mathrm{S}$ method~\cite{Junk:1999kv, Read:2002hq} to incorporate the systematic uncertainties~\cite{Cousins:1991qz}. These limits are presented in Fig.~\ref{fig:70gevxsec} as a function of particle lifetime $\tau$. The two left-hand axes of Fig.~\ref{fig:70gevxsec} are production cross section times branching fraction ($\sigma \times \mathcal{B}$) for top squarks and gluinos, assuming the total visible energy in the decay satisfies either $E_{\Pg} > 120\GeV$ or $E_{\PQt} > 150\GeV$ for the gluino and top squark analyses, respectively. The minimum energy of the SM particle is set by considering the reconstruction efficiency shown in Fig.~\ref{fig:deteff}. Below this energy, the reconstruction efficiency drops off rapidly and we are significantly less sensitive to \PSg{} and \PSQt decays.  By not making a specific neutralino mass hypothesis, we are able to constrain a larger phase space of top squark decays, including the region where the top squarks are off mass-shell. The right-hand axis of Fig.~\ref{fig:70gevxsec} shows the quantity $\sigma \times \mathcal{B} \times\varepsilon_\text{stopping} \times\varepsilon_\text{reco}$, which is more model independent.

\subsection{Limits on gluino and top squark mass}
\begin{figure}[th]
 \centering
   \includegraphics[width=\cmsFigWidth]{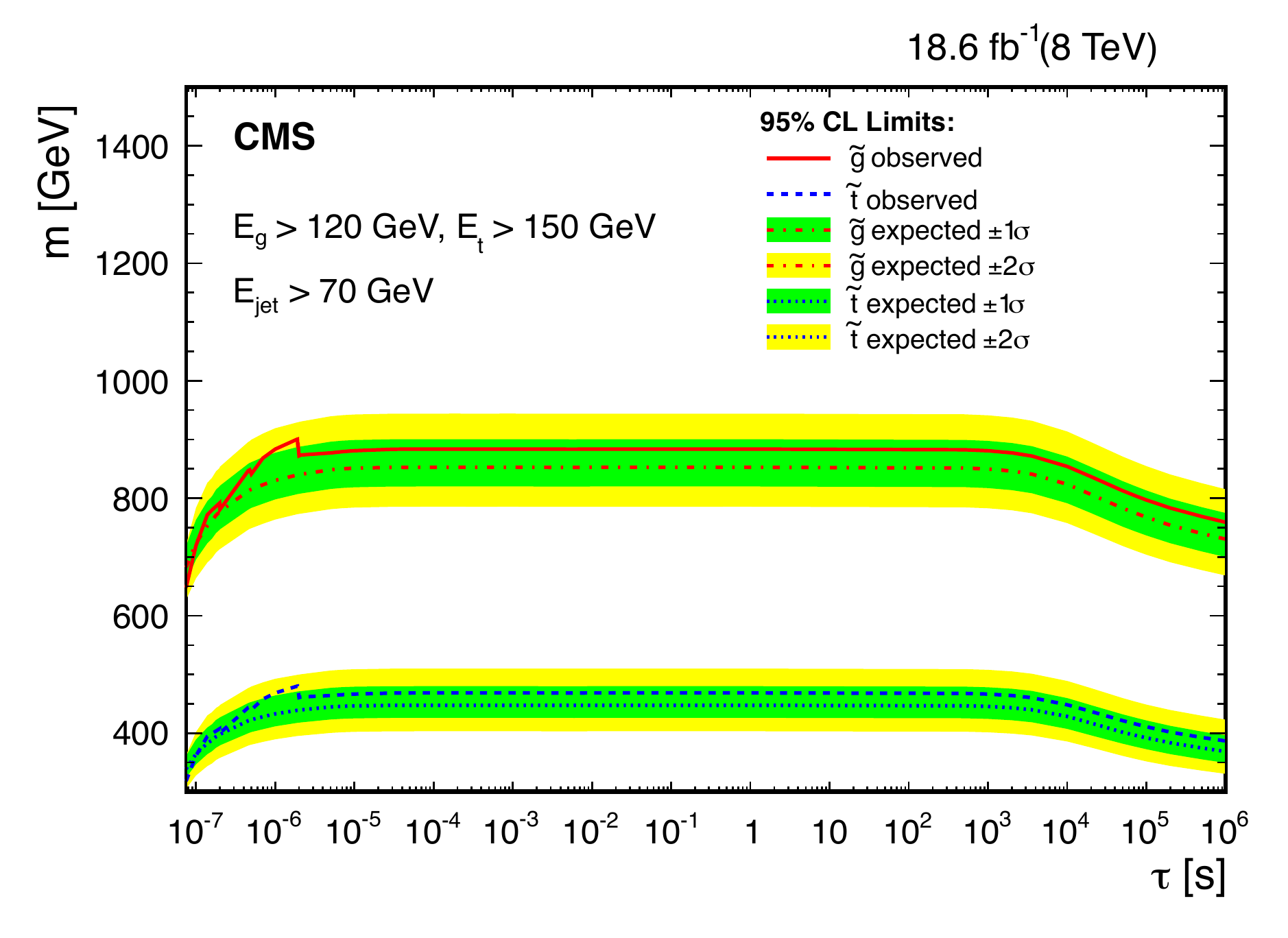}
   \caption{Lower limits at 95\% CL on gluino and top squark mass as a function of particle lifetime, assuming the cloud model of R-hadron interactions and NLO+NLL production cross sections given in Ref.~\cite{Kramer:1456029}. When $E_{\PQt} < m_{\PQt}$, the top quark is allowed to go off the mass shell. }
   \label{fig:70gevmass}
\end{figure}

Figure~\ref{fig:70gevmass} shows the limits on gluino and top squark mass as a function of the particle lifetime. The production cross sections at $\sqrt{s}=8\TeV$ were obtained at next-to-leading order in $\alpha_s$ with next-to-leading-logarithmic soft gluon summation (NLO+NLL) using \textsc{NLL-Fast}~\cite{Beenakker:2011fu}, with the assumption that any other sparticles are decoupled. Assuming $\mathcal{B}(\PSg\to \Pg\PSGcz) = 100\%$ and $\mathcal{B}(\PSQt\to \PQt\PSGcz) = 100\%$,
we are able to exclude $m_{\PSg} < 880\GeV$ and $m_{\PSQt}< 470\GeV$ at 95\% CL for $1\mus < \tau < 1000\unit{s}$ with $E_{\Pg} > 120\GeV$ and $E_{\PQt} > 150\GeV$. Because of the requirements on the minimum energies for the gluon (top quark), these limits do not apply for all neutralino masses, as discussed in the next section.

\subsection{Results for higher energy thresholds}
\label{sec:higherEnergy}
With the selection criteria described previously, we are able to reduce background contamination to acceptable levels. We can, however, be more aggressive with the removal of backgrounds by increasing the offline jet energy threshold ($E_\text{thresh}$). Since $\varepsilon_\text{reco}$ is essentially flat above the minimum energy of $E_{\Pg}$ or $E_\PQt$, and the background falls steeply with energy, we potentially obtain stronger limits on the production cross section by running the analysis with an increased jet energy threshold. This more aggressive method of reducing background was performed for $E_\text{thresh} = 100,$ 150, 200, and 300\GeV. However, as $E_\text{thresh}$ increases, the sensitivity to heavy $\PSGcz$ degrades. If there is a smaller mass splitting between $\PSg (\PSQt)$ and $\PSGcz$, the amount of energy available for the visible decay product is small.

\begin{figure*}[th]
 \centering
   \includegraphics[width=0.48\linewidth]{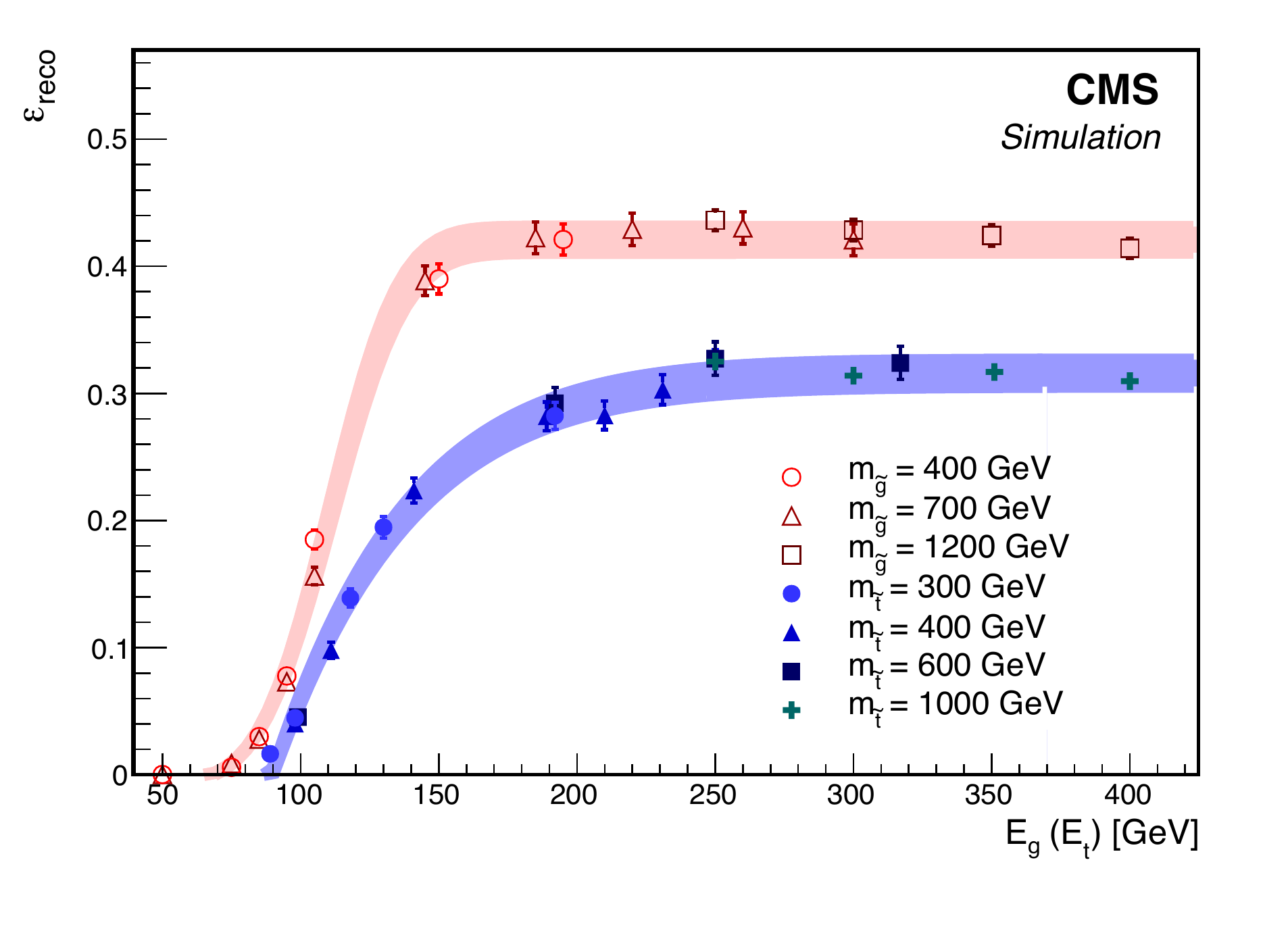} \includegraphics[width=0.48\linewidth]{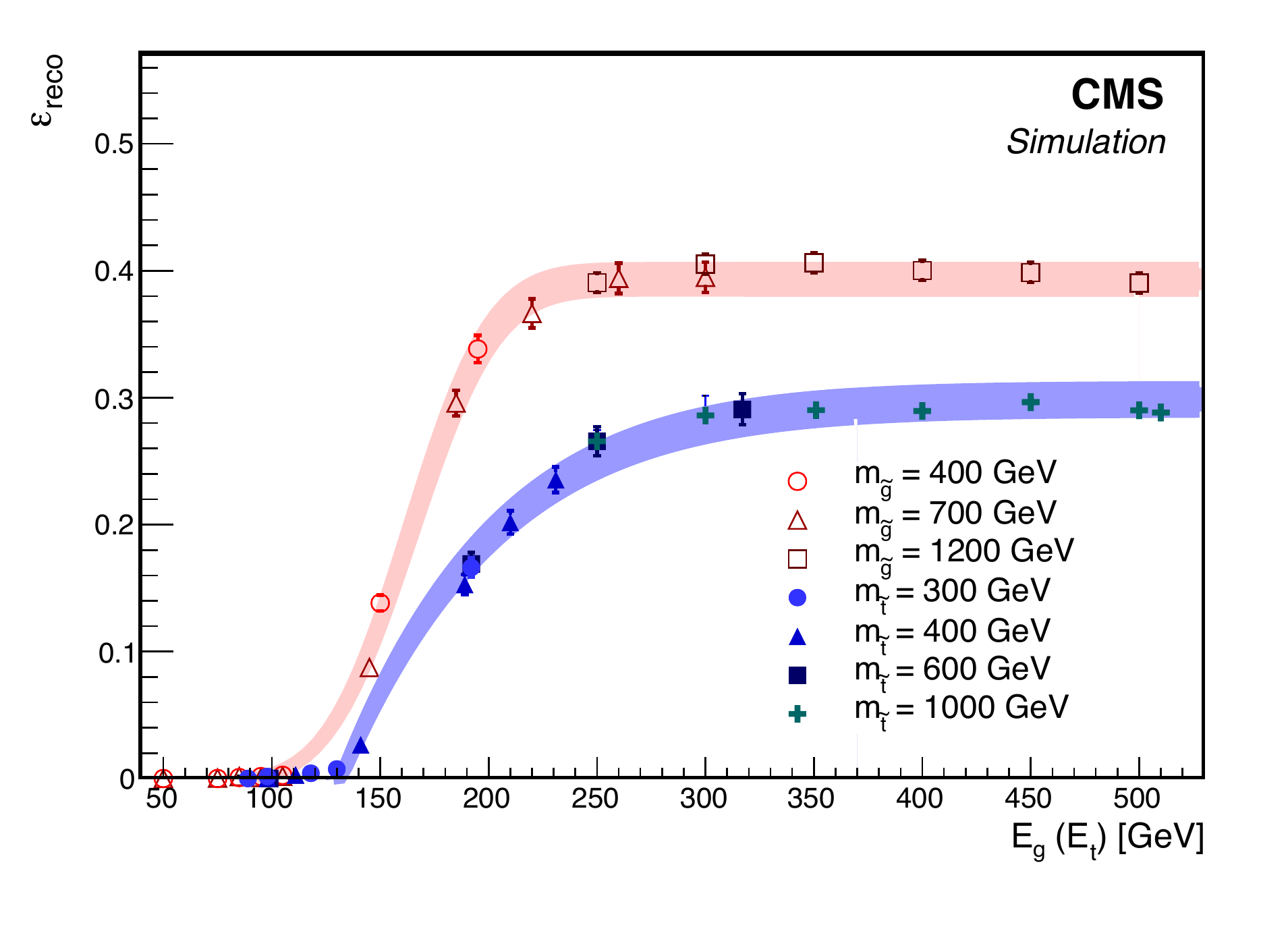}
   \includegraphics[width=0.48\linewidth]{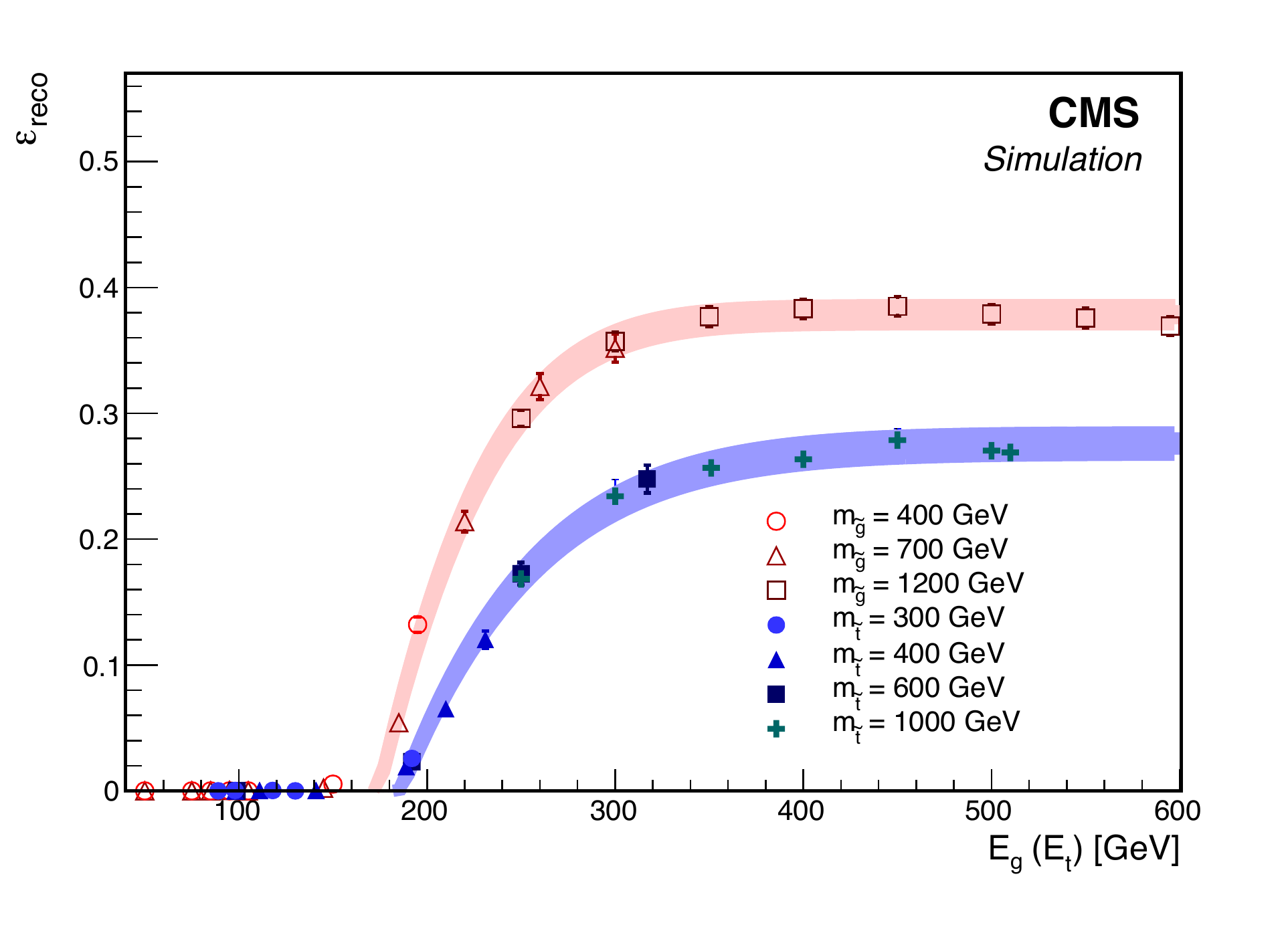} \includegraphics[width=0.48\linewidth]{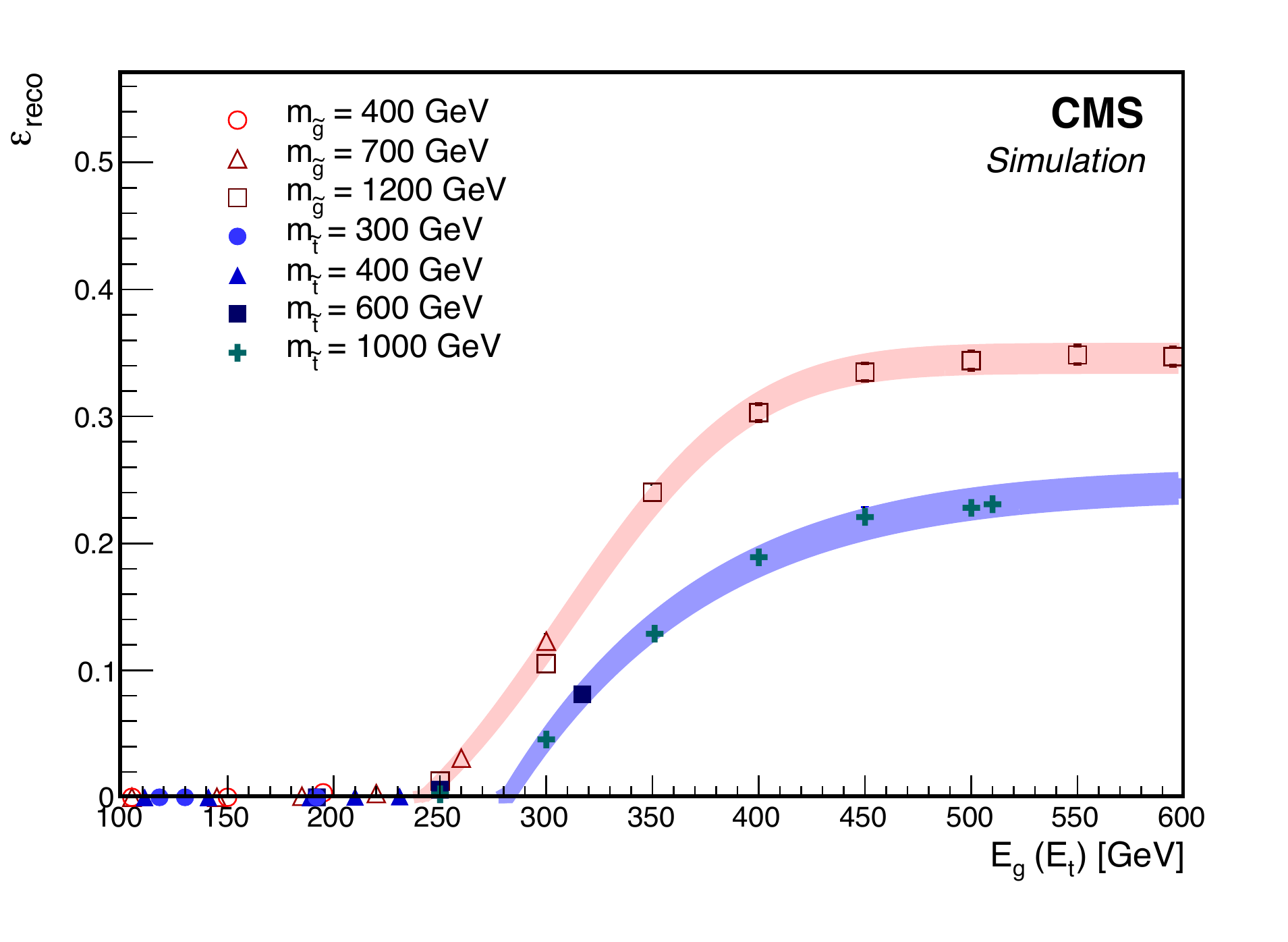}

   \caption{The reconstruction efficiency $\varepsilon_\text{reco}$ for \PSg{} and \PSQt R-hadrons that stopped in the barrel region of the calorimeter as a function of the energy of the SM daughter particle for jet energy thresholds of 100, 150, 200, and 300\GeV.}
   \label{fig:deteffall}
\end{figure*}

To perform the analysis at the higher jet energy thresholds, the threshold is applied to the simulated signal to calculate the minimum energy of $E_{\Pg}$ or $E_\PQt$ and $\varepsilon_\text{reco}$ for each ${\rm E_{ thresh}}$. We repeat the analysis of the 2010 data to estimate the instrumental noise rate at the increased threshold, and then the cosmic and beam halo rates are determined as for the analysis with the 70\GeV jet energy threshold. The resultant contributions of each background source to each signal region are presented in Table~\ref{tab:threshbkg}. Limits on gluino and squark masses for each threshold are presented in Table~\ref{tab:threshresults}. These limits are valid for the minimum value of $E_{\Pg}$ and $E_\PQt$ that we calculate from the turn-on curves shown in Fig.~\ref{fig:deteffall}. These minimum values are listed in Table~\ref{tab:threshresults} for each threshold; they increase with increased ${\rm E_{ thresh}}$ because the turn on plateau for $\varepsilon_\text{reco}$ moves in response to the higher thresholds as seen in Fig.~\ref{fig:deteffall}.

The systematic uncertainties in $\varepsilon_\text{reco}$ and integrated luminosity are unaffected by the increase in the jet energy threshold. However, the systematic uncertainty resulting from the JES does vary somewhat with different ${\rm E_{ thresh}}$. The final JES uncertainty is calculated by measuring the change in $\varepsilon_\text{reco}$ when the jet energy threshold requirement is varied according to the JES systematic uncertainty. Variations in the jet energy requirement have the largest impact for gluon (top) energies close to the threshold, so we perform this calculation on simulated signal samples corresponding to the minimum values of $E_{\Pg}$ ($E_{\PQt}$).

\begin{figure}[th]
 \centering
   \includegraphics[width=\cmsFigWidth]{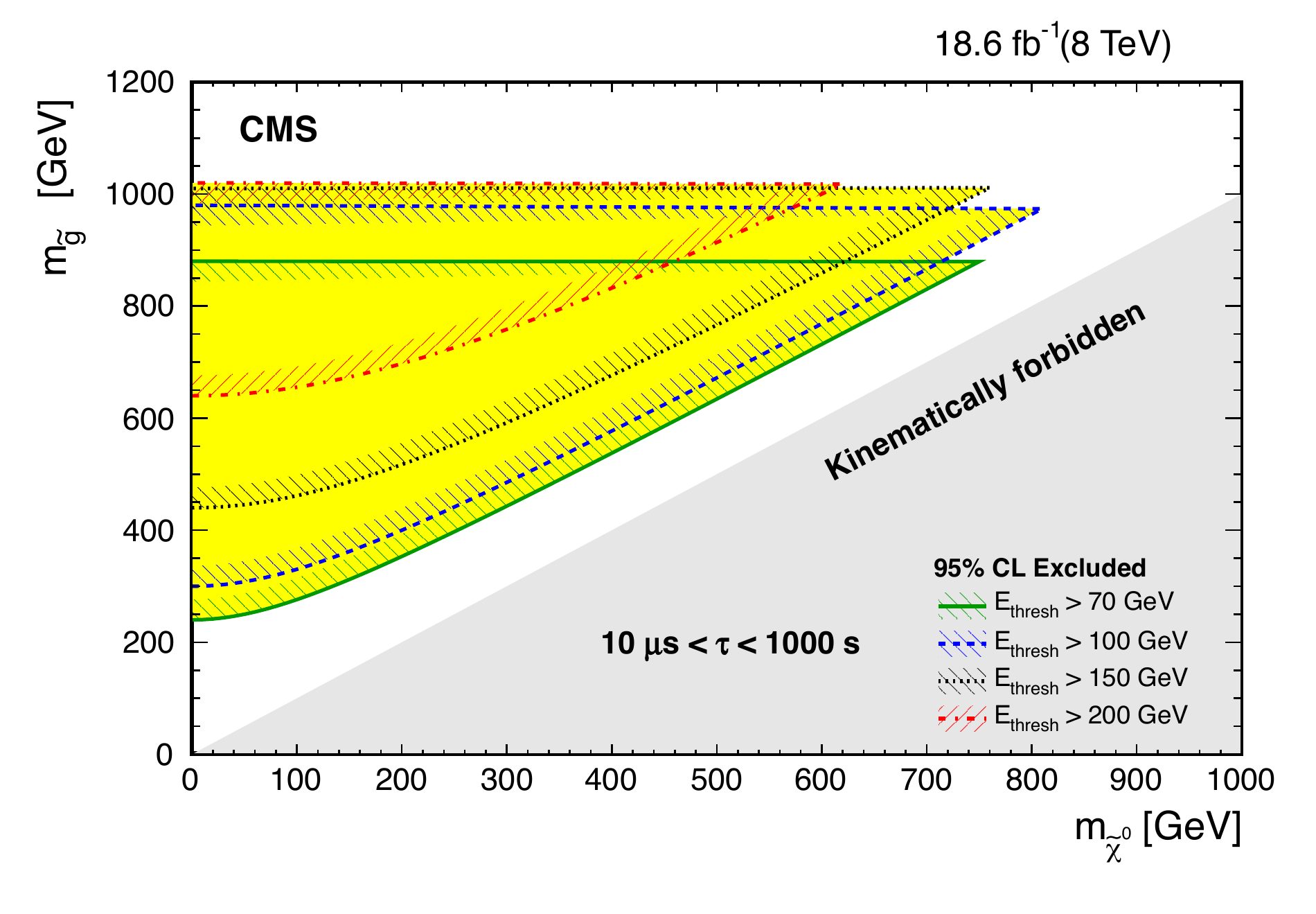}
   \caption{Regions of the $m_{\PSg} - m_{\PSGcz}$ plane excluded by the analysis, valid for $10^{-5}\unit{s}< \tau < 10^3\unit{s}$ using thresholds as indicated in the legend.}
   \label{fig:excluded}
\end{figure}

\begin{figure}[th]
 \centering
   \includegraphics[width=\cmsFigWidth]{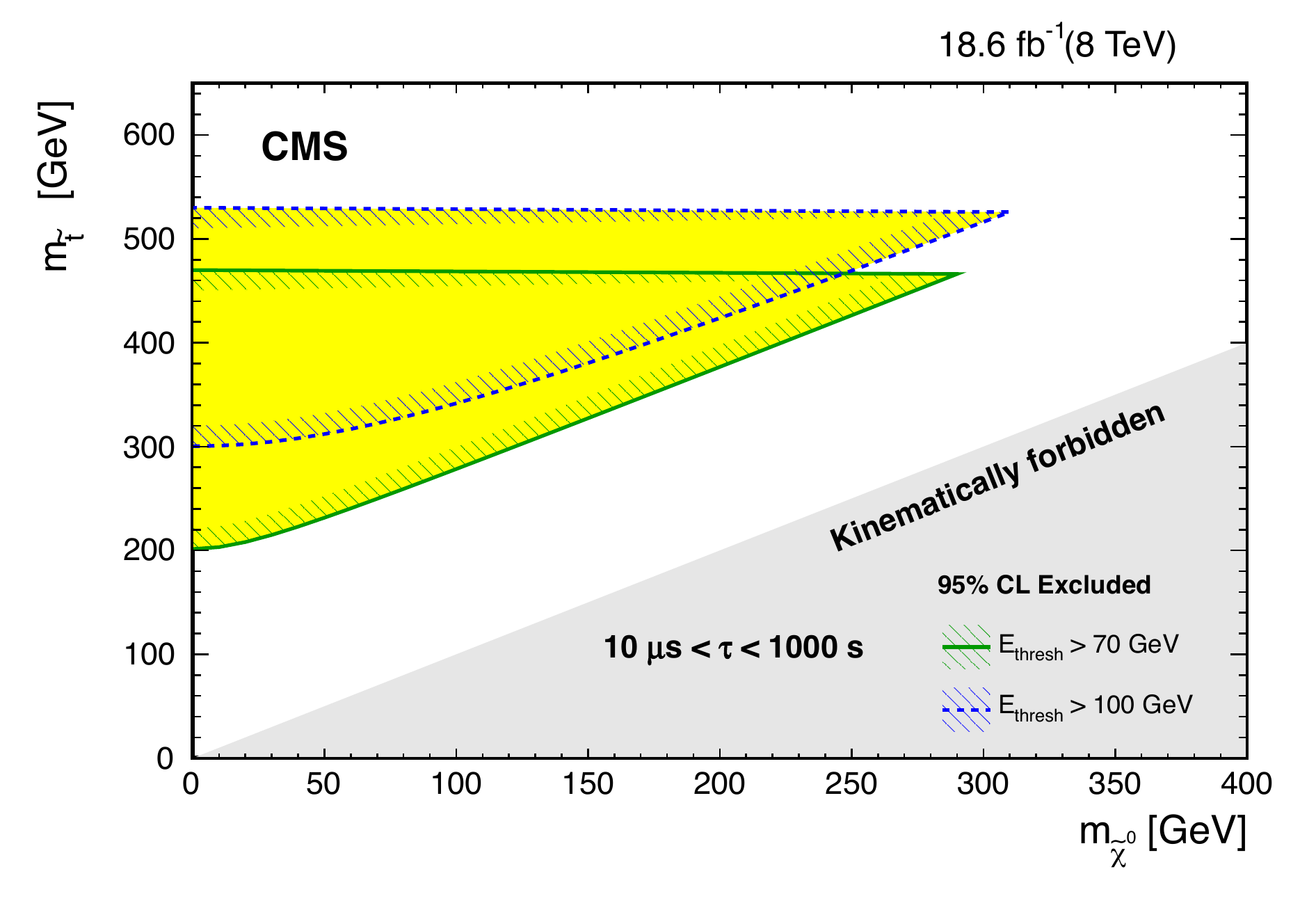}
   \caption{Regions of the $m_{\PSQt} - m_{\PSGcz}$ plane excluded by the analysis, valid for $10^{-5}\unit{s} < \tau < 10^3\unit{s}$ using two energy thresholds as indicated in the legend. The excluded regions only apply to on-shell top quarks.}
   \label{fig:excludedStop}
\end{figure}

{\tolerance=400
As mentioned previously, increasing ${\rm E_{ thresh}}$ affects the masses of $\PSGcz$ that are accessible to the analysis. Figure~\ref{fig:excluded} summarizes how these different jet energy thresholds exclude different regions of the $(m_{\PSg}, m_{\PSGcz})$ phase space. Figure~\ref{fig:excludedStop} does the same for the $(m_{\PSQt}, m_{\PSGcz})$ phase space, though it only applies to on-shell top quark decays because $m_{\PSGcz}$ is unknown when the top goes off mass-shell. It should be noted that the minimum lifetime for the higher threshold limits increases from 1\mus to 10\mus. This decrease in sensitivity to smaller lifetimes is due to the smaller sample size associated with the increased energy requirement.
\par}

\begin{table}[th]
  \topcaption{Background estimates for various energy thresholds.}
  \label{tab:threshbkg}
  \centering
    \renewcommand{\arraystretch}{1.25}
  \begin{tabular}{cccccc} \hline
	$E_\text{thresh}$ (\GeVns)	&$N^\text{bkg}_\text{noise}$	&$N^\text{bkg}_\text{cosmic}$	&$N^\text{bkg}_\text{halo}$	&$N^\text{bkg}_\text{total}$\\ \hline
	
	70		&$0.0^{+2.6}_{-0.0}$	&$5.2 \pm 2.5$		&$8.0\pm 0.4$		&$13.2^{+3.6}_{-2.5}$\\
	100		&$0.0^{+2.0}_{-0.0}$	&$3.1\pm1.2$		&$1.7\pm 0.4$		&$4.9^{+2.4}_{-1.2}$	\\	
	150		&$0.0^{+2.2}_{-0.0}$	&$1.6\pm1.0$		&$0.6\pm 0.1$		&$2.1^{+2.4}_{-1.0}$	\\
	200		&$0.0^{+1.3}_{-0.0}$	&$0.5\pm0.4$		&$0.5\pm 0.1$		&$0.7^{+1.4}_{-0.4}$	\\
	300		&$0.0^{+1.3}_{-0.0}$	&$0.4\pm0.4$		&$0.04\pm 0.02$		&$0.4^{+1.3}_{-0.4}$	\\ \hline
  \end{tabular}

\end{table}

\begin{table*}[thb]
  \caption{Lower limits on gluino and top squark masses obtained from the analyses with varied jet energy thresholds.}
  \label{tab:threshresults}
  \centering
  \renewcommand{\arraystretch}{1.25}
  \begin{tabular}{ccccccc} \hline
	$E_\text{thresh}$ (\GeVns)	&$N_{bkg}$	&$N_{obs}$ 	&$E_{\Pg}^\text{min} (\GeVns)$	&$m_{\PSg}$ limit (\GeVns) &$E_{\PQt}^\text{min} (\GeVns)$	&$m_{\PSQt}$ limit (\GeVns)\\ \hline
	70	&$13.2^{+3.6}_{-2.5}$	&10 	&120	& 880	&150	& 470\\
	100	&$4.9^{+2.4}_{-1.2}$		&1		&150	& 990	&200	& 530\\
	150	&$2.1^{+2.4}_{-1.0}$		&0		&220	& 1010	&300	& 550\\
	200	&$0.7^{+1.4}_{-0.4}$		&0		&320	& 1020	&360	& 550\\
	300	&$0.4^{+1.3}_{-0.4}$		&0		&430	&1020	&470	& 550 \\ \hline
  \end{tabular}

\end{table*}

\section{Summary}
\label{sec:conclusion}
A search has been made for long-lived particles that have stopped in the CMS detector after being produced in 8\TeV pp collisions at the CERN LHC. The subsequent decay of these particles was looked for during gaps between proton bunches in the LHC beams.  In a data set with a peak instantaneous luminosity of $7.5 \times 10^{33}\cm^{-2}\unit{s}^{-1}$, an integrated luminosity of 18.6\fbinv, and a search interval corresponding to 281~hours of trigger livetime, no excess above background is observed. Limits are presented at 95\% CL on gluino and top squark production over 13 orders of magnitude in the mean proper lifetime of the stopped particle. Assuming a cloud model of R-hadron interactions, for $E_{\Pg} > 120\GeV$, and $\mathcal{B}(\PSg\to \Pg\PSGcz) = 100\%$, gluinos with lifetimes from 1\mus to 1000\unit{s} and $m_{\PSg} < 880\GeV$ are excluded. Under similar assumptions, $E_{\PQt} >150\GeV$, and $\mathcal{B}(\PSQt\to \PQt\PSGcz) = 100\%$, long-lived top squarks with lifetimes from 1\mus to 1000\unit{s} and $m_{\PSQt} < 470\GeV$ are excluded. By increasing the jet energy requirement, these mass exclusions increase to $m_{\PSg} \lesssim 1000\GeV$ and $m_{\PSQt} \lesssim  525\GeV$ in a more restricted region of parameter space. In all cases, these exclusions require that $m_{\PSGcz}$ is kinematically consistent with the minimum values of $E_{\Pg}$ and $E_{\PQt}$. These results are the most stringent constraints on stopped particles to date.

\begin{acknowledgments}
We congratulate our colleagues in the CERN accelerator departments for the excellent performance of the LHC and thank the technical and administrative staffs at CERN and at other CMS institutes for their contributions to the success of the CMS effort. In addition, we gratefully acknowledge the computing centers and personnel of the Worldwide LHC Computing Grid for delivering so effectively the computing infrastructure essential to our analyses. Finally, we acknowledge the enduring support for the construction and operation of the LHC and the CMS detector provided by the following funding agencies: BMWFW and FWF (Austria); FNRS and FWO (Belgium); CNPq, CAPES, FAPERJ, and FAPESP (Brazil); MES (Bulgaria); CERN; CAS, MoST, and NSFC (China); COLCIENCIAS (Colombia); MSES and CSF (Croatia); RPF (Cyprus); MoER, ERC IUT and ERDF (Estonia); Academy of Finland, MEC, and HIP (Finland); CEA and CNRS/IN2P3 (France); BMBF, DFG, and HGF (Germany); GSRT (Greece); OTKA and NIH (Hungary); DAE and DST (India); IPM (Iran); SFI (Ireland); INFN (Italy); MSIP and NRF (Republic of Korea); LAS (Lithuania); MOE and UM (Malaysia); CINVESTAV, CONACYT, SEP, and UASLP-FAI (Mexico); MBIE (New Zealand); PAEC (Pakistan); MSHE and NSC (Poland); FCT (Portugal); JINR (Dubna); MON, RosAtom, RAS and RFBR (Russia); MESTD (Serbia); SEIDI and CPAN (Spain); Swiss Funding Agencies (Switzerland); MST (Taipei); ThEPCenter, IPST, STAR and NSTDA (Thailand); TUBITAK and TAEK (Turkey); NASU and SFFR (Ukraine); STFC (United Kingdom); DOE and NSF (USA).

Individuals have received support from the Marie-Curie program and the European Research Council and EPLANET (European Union); the Leventis Foundation; the A. P. Sloan Foundation; the Alexander von Humboldt Foundation; the Belgian Federal Science Policy Office; the Fonds pour la Formation \`a la Recherche dans l'Industrie et dans l'Agriculture (FRIA-Belgium); the Agentschap voor Innovatie door Wetenschap en Technologie (IWT-Belgium); the Ministry of Education, Youth and Sports (MEYS) of the Czech Republic; the Council of Science and Industrial Research, India; the HOMING PLUS program of Foundation for Polish Science, cofinanced from European Union, Regional Development Fund; the Compagnia di San Paolo (Torino); the Consorzio per la Fisica (Trieste); MIUR project 20108T4XTM (Italy); the Thalis and Aristeia programs cofinanced by EU-ESF and the Greek NSRF; and the National Priorities Research Program by Qatar National Research Fund.
\end{acknowledgments}

\bibliography{auto_generated}   

\cleardoublepage \appendix\section{The CMS Collaboration \label{app:collab}}\begin{sloppypar}\hyphenpenalty=5000\widowpenalty=500\clubpenalty=5000\textbf{Yerevan Physics Institute,  Yerevan,  Armenia}\\*[0pt]
V.~Khachatryan, A.M.~Sirunyan, A.~Tumasyan
\vskip\cmsinstskip
\textbf{Institut f\"{u}r Hochenergiephysik der OeAW,  Wien,  Austria}\\*[0pt]
W.~Adam, T.~Bergauer, M.~Dragicevic, J.~Er\"{o}, M.~Friedl, R.~Fr\"{u}hwirth\cmsAuthorMark{1}, V.M.~Ghete, C.~Hartl, N.~H\"{o}rmann, J.~Hrubec, M.~Jeitler\cmsAuthorMark{1}, W.~Kiesenhofer, V.~Kn\"{u}nz, M.~Krammer\cmsAuthorMark{1}, I.~Kr\"{a}tschmer, D.~Liko, I.~Mikulec, D.~Rabady\cmsAuthorMark{2}, B.~Rahbaran, H.~Rohringer, R.~Sch\"{o}fbeck, J.~Strauss, W.~Treberer-Treberspurg, W.~Waltenberger, C.-E.~Wulz\cmsAuthorMark{1}
% ...
\vskip\cmsinstskip
% ...
1:~~Also at Vienna University of Technology, Vienna, Austria\\
2:~~Also at CERN, European Organization for Nuclear Research, Geneva, Switzerland\\
% ...

\end{sloppypar}
\end{document}